\documentclass[12pt,preprint]{aastex}

\newcommand{\be}{\begin{eqnarray}}
\newcommand{\ee}{\end{eqnarray}}

\shorttitle{}
\shortauthors{Pfalzner}

\begin{document}

\title{Angular Momentum Transfer in Star-Discs Encounters: The Case of
Low-Mass Discs}
\author{S. Pfalzner}
\affil {I. Physikalisches Institut, University of Cologne, Germany}

\begin{abstract}
A prerequisite for the formation of stars and planetary systems is that 
angular momentum is transported in some way from the inner regions of the 
accretion disc. Tidal effects may play an important part in this angular 
momentum transport. Here the angular momentum transfer in an star-disc 
encounter is investigated numerically for a variety of encounter
parameters in the case of low mass discs. Although good agreement is found 
with analytical results for the entire disc, the loss {\it inside} the 
disc can be up to an order of magnitude higher than previously assumed. The 
differences in angular 
momentum transport by secondaries on a hyperbolic, parabolic and elliptical 
path are shown, and it is found that a succession of distant encounters 
might be equally, if not more, successful in removing angular momentum than 
single close encounter. 
\end{abstract}


\section{Introduction}

The question of angular momentum transport in accretion discs is a long 
standing problem in the theory of star-formation
(\cite{mestel:qjras65,spitzer:proc68}). The typical observed angular
momentum of the cloud cores from which the stars develop is about three orders
of magnitude larger than the maximum that can be contained in a single star
(\cite{bodenheimer:araa95}). Many different processes have been suggested as
potential candidates for angular momentum transport and a detailed review of 
the history and current state of the angular momentum problem has recently 
been made by \cite{larson:mnras02}. The most favoured processes for
angular momentum transfer are viscous torques 
(\cite{shu:aara87,papaloizou:ara95,stahler:proc00,stone:proc00}),
turbulent effects (\cite{klahr:apj03}), magnetic fields (\cite{balbus:apj02}),
change of orbital motion of multiple star systems
(\cite{mestel:qjras65,mouschovias:apj77,larson:mnras02}) and tidal torques 
within the disc. There are strong indications that transport processes 
alone are not sufficient by themselves to solve the angular momentum problem
(\cite{adams:93,bodenheimer:araa95,stahler:proc00,stone:proc00,gammie:01}),  
and additional mechanisms such as tidal effects combined with gravitational 
torques are thought to play an important role, especially in binary and 
multiple systems (\cite{larson:mnras02}). 

In this paper, tidal effects  for encounters between a disc-surrounded primary 
star and a secondary star  will be studied in detail.
One could argue that collision rates determined from observed number 
densities and velocity dispersions are rather low and that collisions
are too rare to play a major role in angular momentum transport. However,
numerical simulations of the fragmentation of molecular clouds produce many 
examples of interactions between fragments with disk-like structures 
(\cite{bate:mnras02}). This leads to the conclusion that encounters might 
be important in the early epochs of star formation. 

There have been earlier investigations of angular momentum transport in
star-disc encounters employing both analytical and numerical methods, but they 
were either  limited to distant encounters 
(\cite{ostriker:apj94,larwood:97}) or studied different aspects of such 
encounters (\cite{heller:apj95,boffin:mnras98,pfalzner:apj03}).
Since Ostriker's analytical investigation was restricted to linear 
perturbation theory, it was naturally confined to distant parabolic 
encounters only. Ostriker defines the limit where perturbation theory breaks 
down at $r_{peri} >  2 r_{disc}$, whereas \cite{hall:mnras96} set this at 
$r_{peri} >  4 r_{disc}$.  Hall et al. performed restricted three-body 
calculations in an investigation of the angular momentum transfer in close 
encounters
treating prograde and retrograde, close and penetrating as well as coplanar 
and non-coplanar cases at different inclinations. As these simulations are 
computationally time-intensive, they only calculated one single encounter on a 
parabolic orbit for each of these cases. 
\cite{hall:mnras96} also find that energy and angular momentum transfer 
are dominated by material that becomes unbound, the only exception being 
prograde encounters, where the angular momentum transfer is dominated by 
material remaining bound to the primary.
 
It is these  prograde, coplanar encounters that will be investigated here,
in particular how the angular momentum transfer depends on the encounter 
parameters. A systematic study will be presented for low mass discs only, 
where simple N-body simulations suffice, and hydrodynamic effects 
and self-gravity within the disc can be largely neglected 
(\cite{pfalzner:apj03}).
It will be demonstrated that, at least for distant encounters, it is
predominantly the outside of the disc which is involved in the angular 
momentum transport; hydrodynamical effects come into play mainly
at the center of the disc where temperature and density are the highest. 
Future work will address high mass discs, with special emphasis on
the effects of self-gravity on the angular momentum transport. 

It is found that for distant encounters, the numerical results agree with the 
analytical results by \cite{ostriker:apj94}, but that considerable 
differences (up to nearly a factor of 3 in the cases considered) exist for 
close encounters.
First, the dependence of the angular momentum transfer on the 
radial distance from the primary star is more complex than previously 
assumed. Second, the angular momentum transport differs
significantly in hyperbolic, parabolic and elliptical cases.  
Third, it is found that successive encounters may lead to far greater
angular momentum reductions than previously thought.
In the case-study investigated  here, it increased the angular
momentum loss by a factor 1.4.

\section{Numerical method and initial conditions}

The disc surrounding the star is simulated using pseudoparticles as
tracers of the observed dust. These test particles feel the
force of the two stars and vice-versa, thus going beyond restricted 3-body 
simulations. Interactions among the disc
particles themselves are neglected, which is a valid simplification
for the low-mass discs ($m_s \ll m_d$) treated here, as demonstrated 
by \cite{pfalzner:apj03}. The simulations trace the path of the 
test particles using a fifth-order Runge-Kutta-Fehlberg integrator with 
variable timestep. The numerical method is described in detail in 
\cite{pfalzner:apj03}.

The mass of the central star $M_p$ is 1 $M _{\sun}$ in all simulations and
located initially in the centre of the disc at the coordinates 
$(x,y,z)=(0,0,0)$. It is allowed to move freely at later times. The mass
of the secondary star varies between $M_s = 0.1 M _{\sun}$ and 
$1.5 M _{\sun}$. The periastron of the encounters is chosen between
100 and 450 AU. Since the disc surrounding the primary star extends to
$r_d$ = 100 AU, this covers the parameter space from distant to grazing
encounters but excludes disc penetration.\\
\\
The density distribution in the disc given by
\be \label{rho}
     \rho(r,z) = \rho_0(r) \exp \left( - \frac{z^2}{2 \, H(r)^2} \right),
\label{eq:mass}
\ee
where $H(r)$ is the local vertical half-thickness of the disc, which
depends on the temperature, and $\rho_0(r)$ is the mid-plane
density. All simulations presented have used 10 000 simulation particles. 
However, some additional test calculations with 50 000 particles were 
performed (not included in the table). Because no 
real differences to the simulations with
10 000 particles were found, it was concluded that 10 000 particles are 
sufficient to understand the global scaling of angular momentum transfer
with interaction parameters. 
However, it should be pointed out that for more distant encounters, the error 
bar in the angular momentum change is larger for 10 000 particles 
than 50 000 because of the reduced resolution at the outer edge.
 
The disc mass is $m_{disc}$ = 0.001 $M _{\sun}$.
A gap of 10 AU  between the star and the inner edge of the disc has 
been assumed in order to save computer time and to avoid additional complex 
calculation of direct star/disc interactions.

To a first approximation the results can be generalized to larger disc sizes,
since the disc mass is very small and the effects of the disc 
particles on the stars very limited. The same holds for the 
density distribution: 
the results can to converted to an arbitrary density 
profile $\Theta(r)$ by simple multiplication with 
$\Theta(r)/\Theta_{-1}(r) 2\pi r \Delta r$, where $\Delta r$ is the radial 
extent of each annulus of material and $\Theta_{-1}(r)$ the $1/r$-density 
profile used in above simulations.

\section{Angular momentum transfer in encounter}

\subsection{Radial dependence}

\begin{figure}
\epsscale{.85}
\plotone{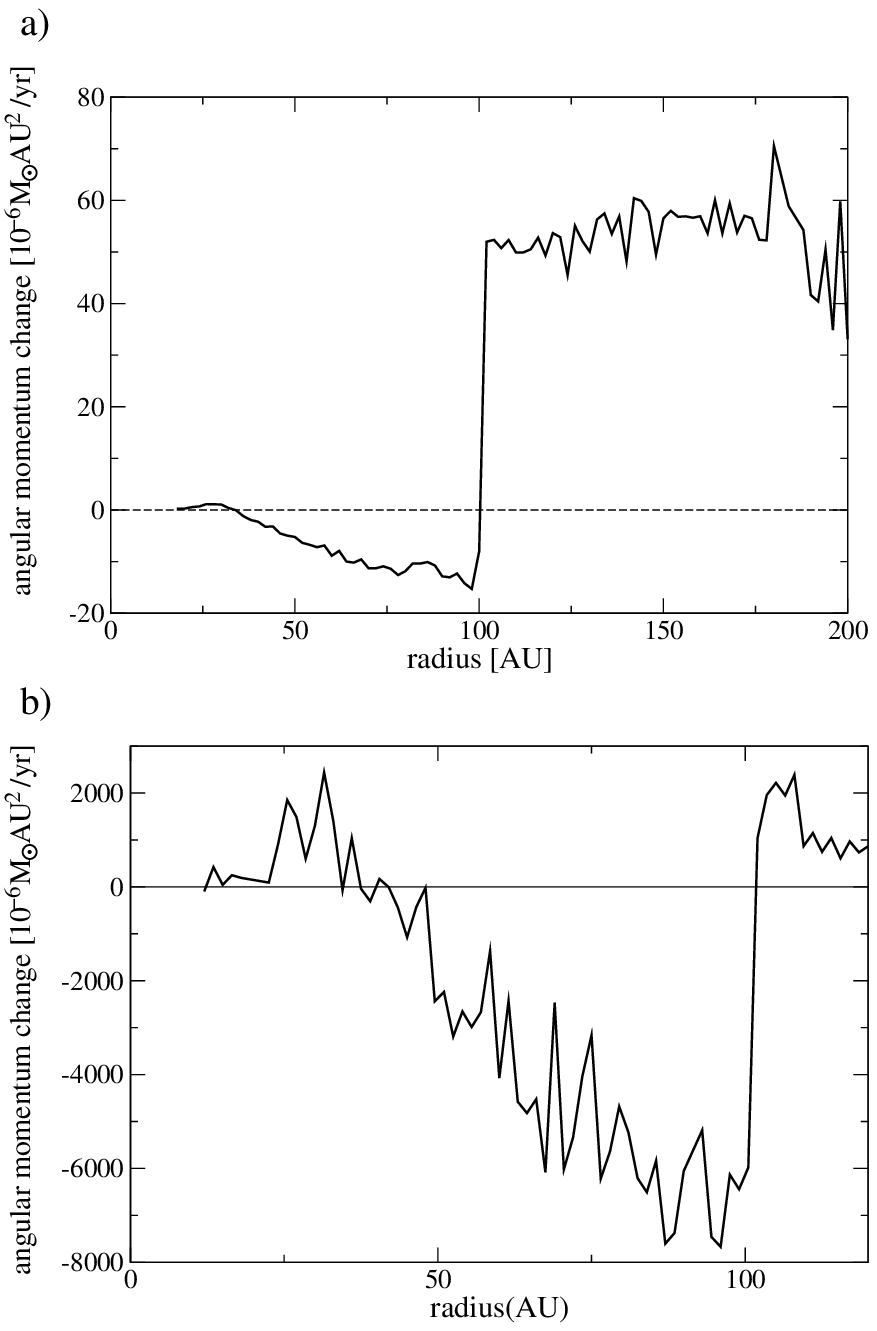}
\caption{Change of the angular momentum  a) per 
particle and b) in total as a function of the radial distance from
the primary star.
\label{fig:ang_rad_simple}}
\end{figure}

Fig. \ref{fig:ang_rad_simple}a) shows a typical example of the change of the 
angular momentum per particle (or per mass unit: these are equivalent in this 
case, since all simulation particles are of equal mass) as a function of the 
distance from the central star in one of our simulations. Outside the 
original disc radius (100AU) only a gain in angular momentum can occur, since 
there were obviously no particles in this region before the encounter. Inside 
this initial radius there is a broad region where each particle loses angular 
momentum and still further inside is a 
small region where angular momentum is gained (but to a much lesser extent). 
Close to the central star (not shown here) the angular momentum remains 
unaltered. Apparently the inner angular momentum gain has not been seen 
previously: its physical origin remains yet to be determined. 

The change of angular momentum of the disc is usually measured by summing 
over all particles still bound to the central star after the encounter.
Although the gain of angular momentum per particle outside the initial
disc radius is larger than the loss inside, the total amounts to a loss of 
angular momentum in the disc, since there are many more particles inside the 
disc than outside (see Fig.  \ref{fig:ang_rad_simple}b)). 

Summing over all bound particles - as is usually done - might therefore not be 
appropriate, given that the problem at hand is how to lose angular 
momentum from the central region of the disc. The particles outside the 
original disc do not really figure in this process, apart from compensating 
to some degree for the loss in the more central areas of the disc by their 
angular momentum gain.
Simulations show that the largest decrease of angular momentum occurs
integrating up the original disc size (in this case 100 AU), for smaller or 
larger radii the loss is less marked. Therefore in this paper the change of 
angular momentum {\it within the original disc size of 100AU} will be taken as 
the figure of merit; the 
change in the entire disc will only be used when comparing to previous work.

\begin{figure}
\plotone{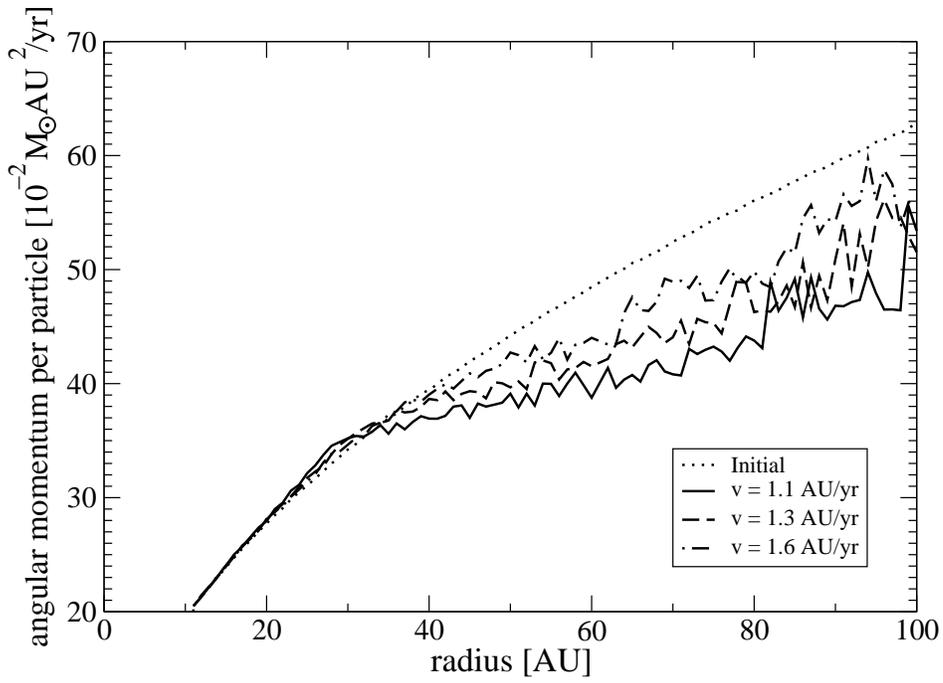}
\caption{Radial dependence of the angular momentum per particles
initially(dotted line) and after an encounter with different velocities of 
the secondary star, but with the same periastron and mass.   
\label{fig:depend1}}
\end{figure}

The encounter parameters mass and periastron significantly change the part 
of the disc which is affect most in terms of 
angular momentum change. A higher mass or closer periastron of the
secondary results in effecting regions further inside the disc. 
By contrast, altering the velocity changes the amount rather than the
location of angular momentum transfer,
(see Fig. \ref{fig:depend1}).

\subsection{Temporal development}

\begin{figure}
\plotone{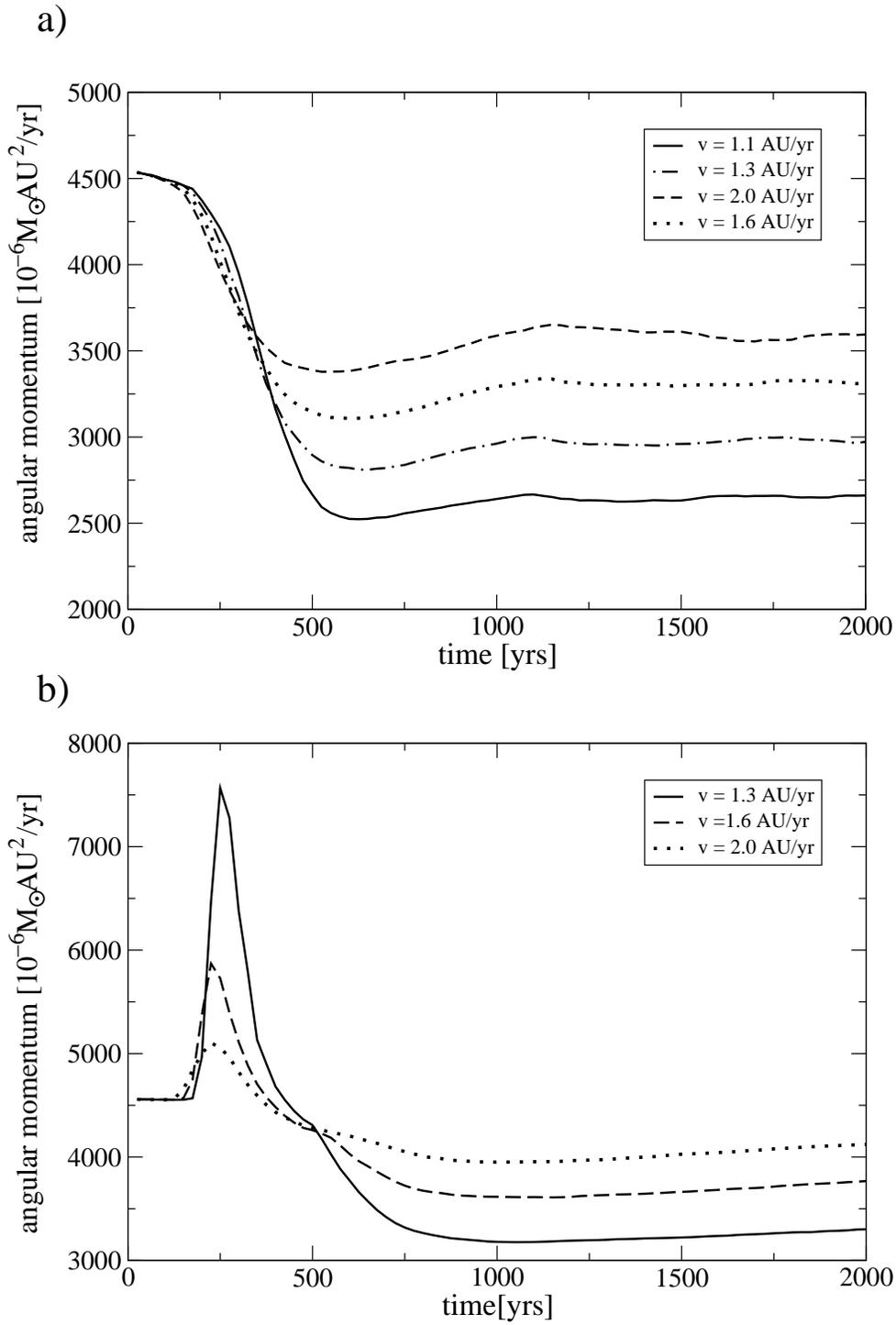}
\caption{Angular momentum a) within the original disc size of 100 AU
and b) of all particles bound to the primary star as a function of time.
\label{fig:ang_temp}}
\end{figure}

The temporal development of the angular momentum during the encounter can be 
seen in Fig \ref{fig:ang_temp} - a) shows this for the angular momentum 
contained within the original 100AU sized disc and b) includes all particles
still bound to the central star. For the latter the angular momentum 
initially increases due to the particles accelerated out of the disc and 
then drops when these particles become unbound.

Turning now to Fig \ref{fig:ang_temp} a) it can be seen that
the angular momentum within the original disc radius decreases, reaches a 
minimum and thereafter settles at a nearly constant value somewhat 
lower 
than the initial angular momentum. It is this difference between the
angular momentum before and after the encounter which will now be considered
in more detail.

\subsection{Dependence on encounter parameters}

First the simulation results are compared with the analytical result 
of Ostriker developed for distant parabolic encounters\cite{ostriker:apj94}. 
This result is complex and requires numerical integration, so that
only values for the case $m_1$ = $m_2$ = 1 $M_\sun$ and $m_d=0.5 M_\sun $ 
with a surface mass distribution $\sim r^{-1}$ are available for comparison. 
If we use such a mass distribution in a simulation it is usually unstable for 
such a high disc mass,
and the steady state solution tends to differ significantly from the
distribution used in the analytical calculation. Therefore, we use a lower 
disc mass and balance it out with a higher central mass. For distant 
encounters the actual path of the secondary is not much altered by the 
extended mass around the primary. In addition,
Ostriker presented the angular momentum change averaged over all
possible encounter angles between the disc and the perturber path. Here the
Ostriker results are modified for coplanar encounters only which gives about 
a factor of 3 higher angular momentum transfer.In these simulations the 
entire extended disc was included, so as to match 
Ostriker's integration over wave numbers. 

\begin{figure}
\plotone{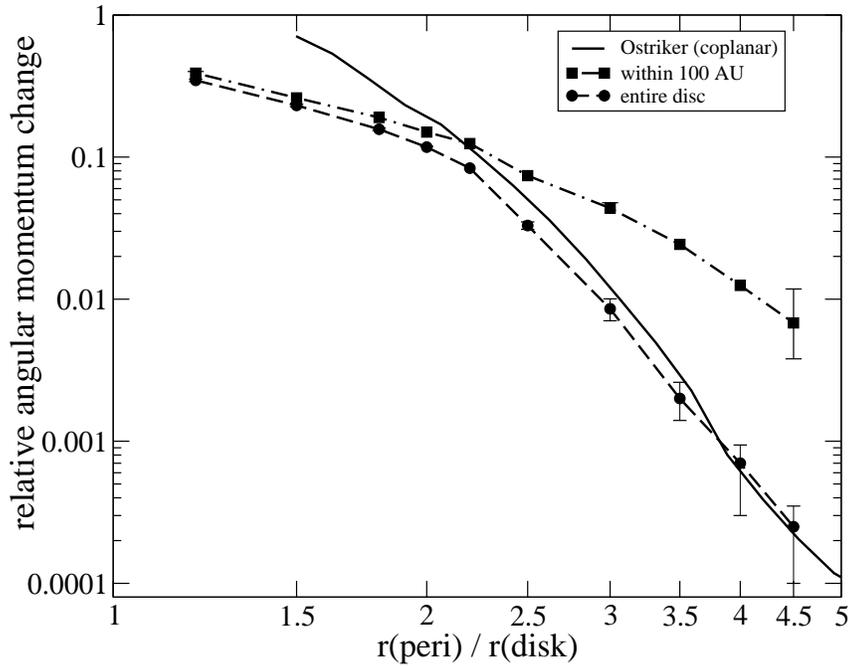}
\caption{Comparison of the analytical results (solid line) 
of relative angular momentum change obtained by
Ostriker(1994) with simulation results for
parabolic coplanar encounters for the entire disc (dashed line) and within the
original disc radius of 100 AU(dash-dotted line).
\label{fig:ostriker}}
\end{figure}

The comparison between the analytical results by Ostriker and the simulation 
results in Fig.\ref{fig:ostriker} show very good agreement for encounters 
more distant than $r_{peri} > 4 r_d$. In this limit one can approximate the 
angular momentum change in the entire disc by simplifying Ostriker's result 
for low mass discs to:
\be \Delta J \sim \pi^2 \frac{4GM_s^2}{5M_p}  \sigma(r_d) 
    \exp \left[ C \frac{r_{peri}}{r_d} \right] \ee
where $C=9 \cdot 2^{5/2}/\Omega(r_d)$. The dependence on $\sigma(r_d)$ and
$\Omega(r_d)$ directly shows its limitations to processes where  
the outer regions of the disc are predominant. For closer encounters this 
relation no longer holds. 

For  $r_{peri} > 2 r_d$ the general trend between the analytical approximation
and the full numerical treatment is similar, but a quantitative relative 
difference of up to 30 per cent is found. For $r_{peri} < 2 r_d$ the results 
deviate considerably, whereby the analytical 
results overestimate the angular momentum transport by up to a factor of 3.

As  argued previously when considering angular momentum loss, it seems more
appropriate just to consider the area of the original disc before
the encounter. The relative angular momentum loss within this region is
much higher, as illustrated by the dash-dotted line in Fig.\ref{fig:ostriker}. 
The relative difference is more pronounced for distant encounters, in which
case the loss of angular momentum could be underestimated by a factor of 10
or more.
For distant encounters the relation $ \Delta J \sim \exp 
\left[ C_2 \frac{r_{peri}}{r_d} \right]$ should still hold, but the actual 
dependencies
of $C_2$ still need to be determined. 

\begin{figure}
\epsscale{.60}
\plotone{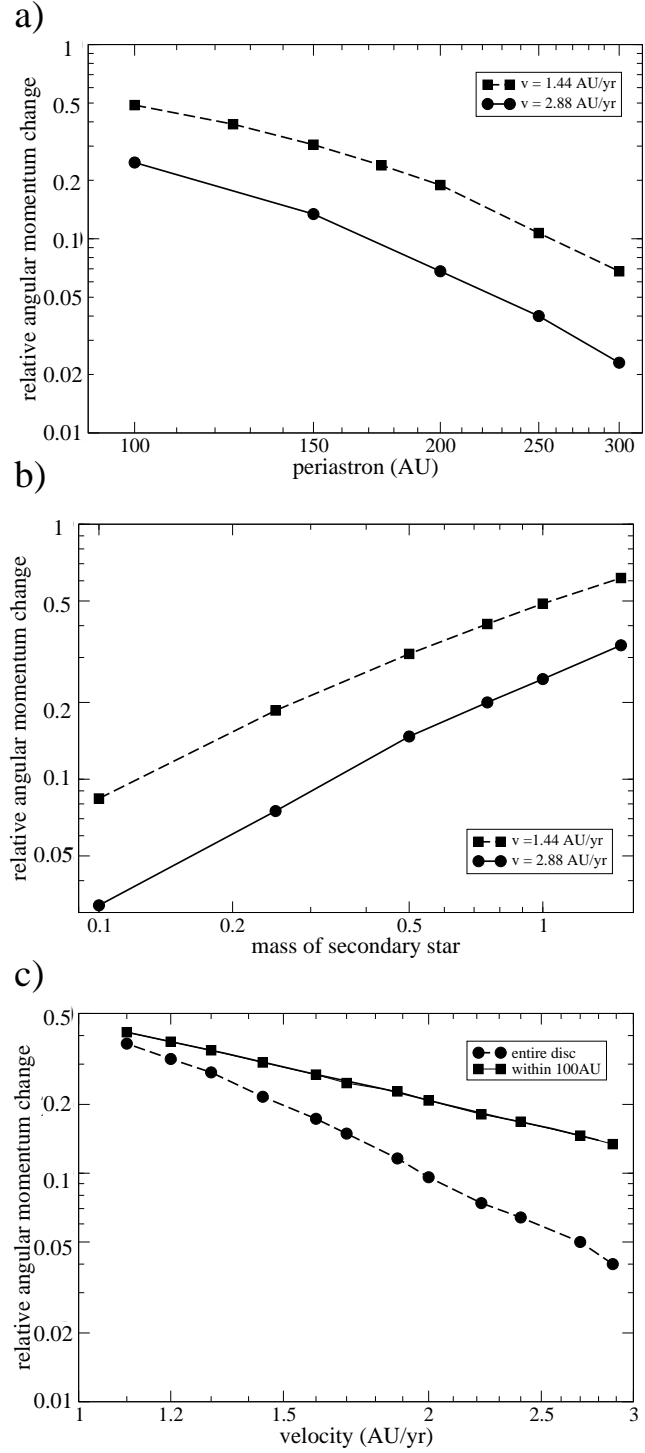}
\caption{The relative angular momentum loss through the encounter
 as a function of a) periastron, b) the mass of the secondary and c) the
relative velocity between the two stars. 
\label{fig:depend}}
\end{figure}

In order to determine some general trends in angular momentum transport in 
{\it close} star-disc encounters, over 60 simulations were performed for
various masses, velocities and periastrons of the secondary star. This is 
the first (although still limited) N-body parameter study of the angular 
momentum transfer in such encounters. The results of these simulations are 
summarized in Fig. \ref{fig:depend} and Table \ref{tab}. Obviously the angular momentum 
change will be greater if the interaction is stronger -  
meaning an encounter with i) a higher mass or ii) a lower relative velocity 
of the secondary or iii) a closer periastron. All these encounters were on 
hyperbolic
orbits,  and therefore the interaction time was relatively short.
Nevertheless,  up to 40 per cent loss of angular momentum (within 100 AU)
was measured for the strongest interactions.

It can be seen that a power law dependence $v^{-a}$ on the velocity of the 
secondary star can be expected ($a=1.168 \pm 0.03$ for original disc and
($a=2.45 \pm 0.05$), but more complex dependencies on the mass and 
periastron of the secondary are found - Fig.\ref{fig:depend}.

Fig.\ref{fig:depend}c) also shows a comparison between the angular 
momentum loss within $r_{disc}$=100AU and the entire disc, as discussed 
earlier. 
It can be seen that the relative difference between these curves grows with 
decreasing encounter strength: for weak encounters, the transferred momentum 
can be underestimated by a factor 10 or more.

\subsection{Orbit type of secondary star}

\begin{figure}
\plotone{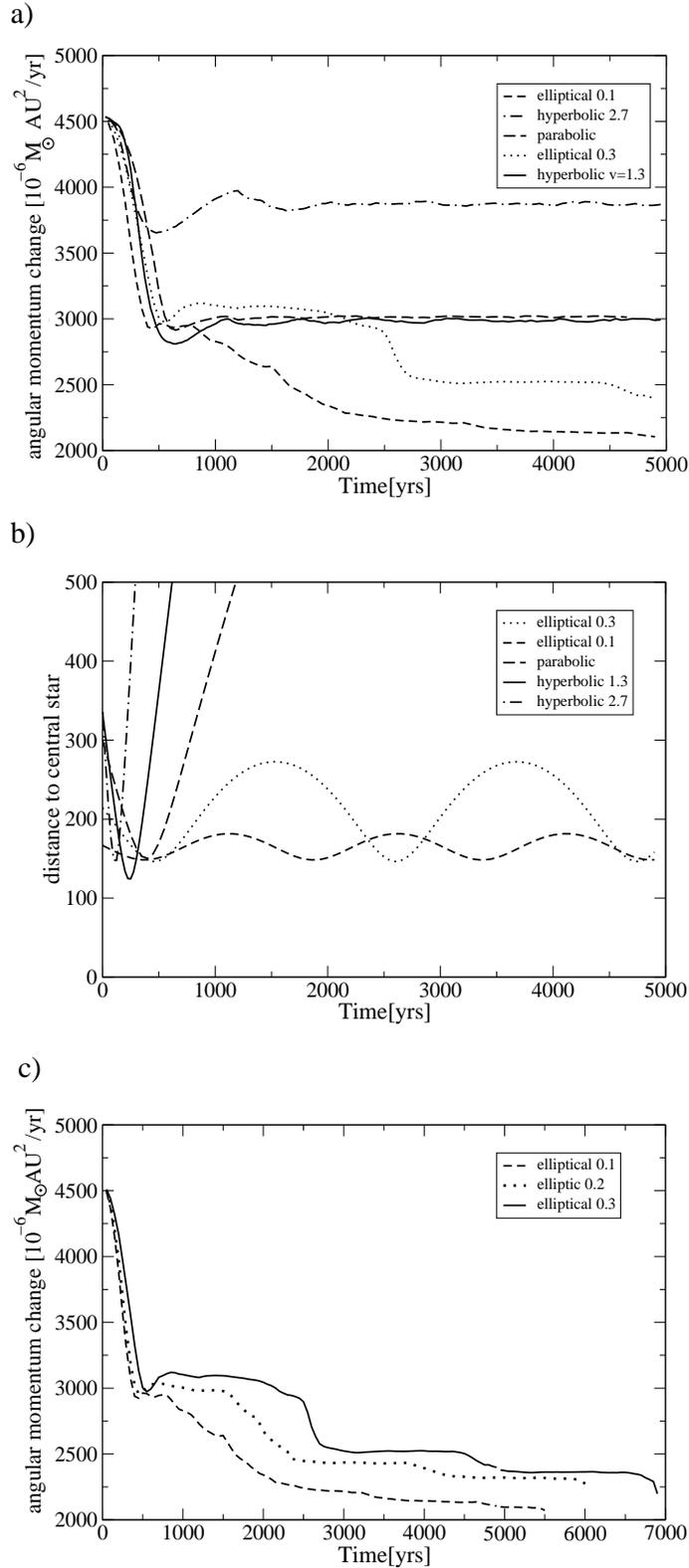}
\caption{a) shows a comparison of the temporal development of the angular 
momentum in elliptic, parabolic and hyperbolic encounters, b) shows the
distance of the secondary to the central star and c) shows the longer term
development just for elliptic encounters of different ellipticity.
\label{fig:comp}}
\end{figure}

It has been argued that as binary or multiple systems form, a much larger
loss of angular momentum can be expected in the disc - \cite{larson:mnras02}. 
In Fig. \ref{fig:comp} 
the temporal development of the angular momentum change 
for different hyperbolic, parabolic and elliptic orbits are compared. 
It seems that for all cases - apart from the hyperbolic cases with 
large $\epsilon$ - the immediate effect of the encounter is
a drop in angular momentum by a fixed amount(small differences in this first
minimum result from slight deviations in the periastron). 

For strongly hyperbolic encounters with large $\epsilon$, the interaction 
time is too short to change the angular momentum to its full extent.
The angular momentum drops to a minimum, increases slightly, 
oscillates for some time, and eventually settles down to a fairly constant 
value. The drop is obvious and the slight increase and following oscillations 
can be explained by the changed orbits of the particles in the spiral arm induced by the encounter
( see \cite{pfalzner:apj03}) - these particles either go on elliptical orbits
or become unbound. So it is either particles still on their way out of
the disc that become unbound later, or particles on elliptical
orbits which cross the radius of 100AU which then are only
considered for the angular momentum change at certain time intervals. 
In a viscous disc, these orbits would be recirculized after some time.

Returning to the other cases, we see that the behaviour of the parabolic and 
less hyperbolic case are very similar.The real difference occurs for the 
elliptical cases - here the ellipticity $\epsilon$ strongly
influences the development as expected.  
Comparing Fig. \ref{fig:comp}a) and b) it can be seen that in the case
$\epsilon$=0.3, renewed loss of angular momentum occurs whenever the secondary
is at the point of closest approach. In the  $\epsilon$=0.1 case, the
link of the angular momentum loss to the periastron is much less distinct.
Here the angular momentum loss seem to be initially much stronger and perhaps
more linked to local effects like interaction with tidal tails. After a
relatively short time the angular momentum decreases only very slowly.

\begin{figure}
\plotone{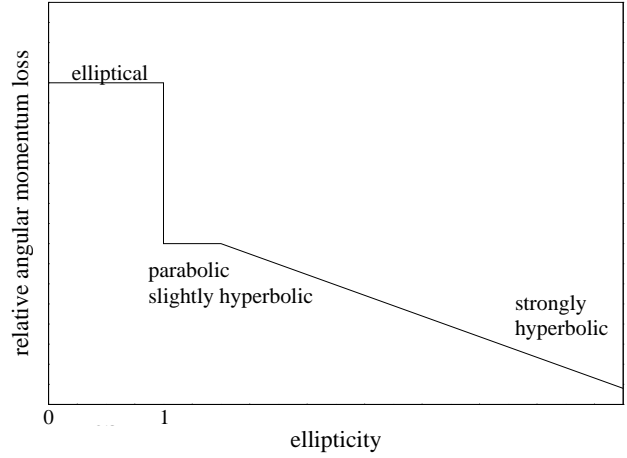}
\caption{Schematic picture of the different regimes of angular momentum loss
for a fixed periastron distance.
\label{fig:regime}}
\end{figure}

There are strong indications in the simulations that for long times the 
angular momentum loss with the secondary on an elliptical orbit tends to the 
same value for all ellipticities, provided the periastron is the same.
For larger ellipticity it just takes longer until it reaches this maximum 
transferable angular momentum. Unfortunately we cannot prove this point yet, 
because
the system becomes numerically unstable after about 10 000yrs. (Elliptical 
orbits need higher order integrators to remain stable, but applying this to all
simulation particles would increase the simulation time considerably.)
However, it is quite likely that a similar dependence for the angular 
momentum transport like that sketched in Fig.\ref{fig:regime} would 
also be found in this case.

The mass and periastron of the secondary determines how far inside the
disc the change of the angular momentum is felt, whereas the velocity
determines how much of the maximum possible angular momentum loss
can actually be achieved within the interaction time.
The parameter space for angular momentum transport in such encounters
could therefore be reduced to two values for each periastron and mass of the 
secondary -
one for the elliptical case and one for the parabolic case. All hyperbolic
encounters can be regarded as to short to fully extract the maximum possible 
angular momentum.

\subsection{Repeated encounters}
\begin{figure}
\plotone{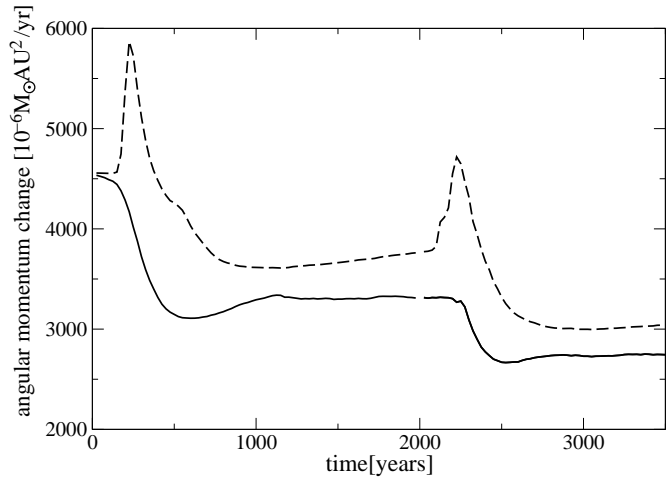}
\caption{Temporal evolution of the angular momentum after a first encounter
at approximately 150yrs and a second encounter with the
same interaction parameters occurring after 2000 years.
In both cases the secondary star has 1 solar mass, the velocity is 1.6AU/yr
and the periastron 150AU.
The dotted line indicates the angular momentum in the entire disc, whereas
the solid line shows that within the original disc radius of 100AU.
\label{fig:double}}
\end{figure}

It is often argued that a second (or consecutive) encounter(s) would not lead 
to any significant additional momentum transport because the outer edge of the
disc is truncated by the first encounter - the part that is mainly involved 
in the interaction. By contrast, the present simulations show 
that a second encounter can indeed lead to a significant angular momentum 
transport( - Fig. \ref{fig:double}). Here, the disc after the encounter is 
exposed to an identical encounter at t = 2000yrs. The absolute angular 
momentum loss is somewhat less than before leading to a total angular momentum
loss of 1.4 times that of the single encounter. However, the relative angular 
momentum loss in this case is even {\it higher} than in the first 
encounter (32 percent vs. 27 per cent in the first encounter).
This result suggests multiple distant encounters should be reconsidered 
as candidate scenarios for significant loss of angular momentum in
the disc.

\section{Discussion and Conclusion}

In this paper the angular momentum loss in the central regions of an 
accretion disc induced by encounters was investigated. A detailed 
study of the dependence on the specific interaction parameters was performed 
by  simulating over 60 different encounter situations. Good agreement was 
found with analytical calculations (Ostriker(1994)) for distant 
parabolic encounters ($r_{peri} > 4 r_d$). However, for close encounters 
($r_{peri} > 2 r_d$) significant deviations are found. In this regime the 
analytical calculations overestimate the angular momentum loss by 
up to a factor 3.

Comparing hyperbolic, parabolic and elliptical systems, it was found that
the mass and periastron of the secondary determine the location of the 
angular momentum transfer within the disc. In hyperbolic encounters, the 
velocity has no influence on the location, but instead determines the 
interaction time and therefore the percentage of the maximum angular 
momentum loss that is actually incurred  during the encounter.
There are strong indications that the angular momentum loss is actually
the same for different elliptical orbits with the same periastron, it just
takes much longer for systems with highly elliptical orbits to reach this 
state.

Generally, the angular momentum loss in the inner regions (which is 
the most relevant to the angular momentum problem) is 
underestimated if the entire disc is included in the calculation. The 
relative difference is most significant for distant encounters, where up 
to a factor of 15 at a periastron of 450 AU was found. 

This implies that a succession of distant 
encounters might well be able to transport a higher amount of angular 
momentum outwards than previously thought. This result is supported by 
the finding that successive encounters can achieve the same (if not more)
relative angular momentum loss than in the first encounter.

To answer this question quantitatively, one would need the probability for 
repeated encounters in clusters of high stellar density (typically 
10$^4$ stars pc$^{-3}$) which to our knowledge has not been investigated so 
far. However, \cite{bonnell:mnras01} and \cite{scally:mnras01} found that 5 to
10 per cent of stars in such an environment undergo a single encounter of
100 AU and closer within the first 2-3 Myr.
For longer time scales ($>$ 10$^7$ yr) their results differ:   
\cite{scally:mnras01} find under 30 per cent of stars undergo such
close collisions, whereas  \cite{bonnell:mnras01} conclude that nearly all
stars have experienced such a close encounter. Whichever scenario is right,
close encounters are clearly {\it not} a rare event in such environments.
Bearing in mind, that distant encounters will be more likely,  
possible loss of angular momentum may therefore be significant.

For close encounters nearly 50 per cent of the angular momentum can be 
removed from within the original disc radius. Although
the transport of angular momentum might not be predominantly
due to encounters, they may nevertheless play an important part.

Finally, it was demonstrated that for close encounters, the area
affected by angular momentum loss can reach far inside the disc
(for the non-penetrating encounters considered here down to 20 AU).
The actual increase in angular momentum near the center occurs where
the disc density has increased most in the encounter. Here,
the density gradient is increased and the effectiveness
of viscous transport later on in the time evolution should be
considerably enhanced.

In summary, encounters appear to have a two-fold effect on angular momentum:
They lead to a considerable angular momentum transport
themselves even more so when a bound system is produced, and 
they probably increase the efficiency of viscous angular momentum
transport near the disc center.

\begin{thebibliography}{alpha}

\bibitem[Adams\&Lin(1993)]{adams:93}
Adams F.C., Lin, D.N.C., in {\em Protostars and Planets III}, ed. Levy, E.H.
Lunine J.I. Univ. of Arizona Press, Tuscon, p.721, 1993.

\bibitem[Balbus\&Hawley(2002)]{balbus:apj02}
Balbus, S.A. and Hawley, J.F.,
{\apj}, 573:749, 2002.

\bibitem[Bate et al.(2002)]{bate:mnras02}
Bate, M.R., Bonnell, I.A., Bromm, V.,
{\em MNRAS}, 336:705, 2003.

\bibitem[Bodenheimer(1995)]{bodenheimer:araa95}
Bodenheimer P., {\em ARAA}, 33:199, 1995.

\bibitem[Boffin et al.(1998)]{boffin:mnras98}
Boffin, H.M.J., Watkins, S.J., Bhattal, A.S.,
		Francis, N., Whitworth, A.P. {\em MNRAS}, 300:1189, 1986.

\bibitem[Bonnell et al.(2001)]{bonnell:mnras01}
Bonnell, I.A., Smith, K.W., Davies, M.B., Horne, K.,
{\mnras}, 322:859, 2001

\bibitem[Gammie(2001)]{gammie:01}
Gammie, C.F.,
{\em ApJ}, 553:174, 2001.

\bibitem[Hall et al.(1996)]{hall:mnras96}
Hall, S.M., Clarke, C.J., Pringle, J.E.
{\em MNRAS}, 278:303, 1996.

\bibitem[Hall(1997)]{hall:mnras97}
Hall, S.M.,
{\em MNRAS}, 287:148, 1997.

\bibitem[Heller(1995)]{heller:apj95}
Heller, C.A.,
{\em ApJ}, 455:252, 1995.

\bibitem[Klahr \& Bodenheimer(2003)]{klahr:apj03}
Klahr, H.~H. and Bodenheimer, P., {\apj}, 582:869, 2003.

\bibitem[Larwood(1997)]{larwood:97}
Nelson, R.P., Papaloizou, J.C.B., Larwood, J.D., Terquem, C.,
In {\em Accretion Disks - New
  Aspects, Proceedings of the EARA Workshop}, ed. H. Meyer-Hofmeister, 182, 1997, Heidelberg,
  Springer.

\bibitem[Larwood \& Kalas(2001)]{larwood:01}
Larwood, J.D., Kalas, P.G.,
  {\em MNRAS}, 323:402, 2001.

\bibitem[Larson(2002)]{larson:mnras02}
Larson, R. B., {\em MNRAS}, 332:155, 2002.

\bibitem[Mestel(1965)]{mestel:qjras65}
Mestel, L. {\em QJRAS} 6:161, 1965.

\bibitem[Mouschovias(1977)]{mouschovias:apj77}
Mouschovias, T. Ch., {\em ApJ}, 211:147, 1977.

\bibitem[Ostriker(1994)]{ostriker:apj94}
Ostriker, C.E., {\em ApJ}, 424:292, 1994.

\bibitem[Papaloizou \& Lin(1995)]{papaloizou:ara95}
Papaloizou, J. C. B., Lin, D. N. C.
  {\em ARAA}, 33:505, 1995.

\bibitem[Pfalzner(2003)]{pfalzner:apj03} 
Pfalzner, S.  {\em ApJ}, 592:986 (2003).

\bibitem[Scally \& Clarke(2001)]{scally:mnras01}
Scally, A. and Clarke, C.,  {\em MNRAS}, 325:449, 2001.

\bibitem[Shu et al(1987)]{shu:aara87}
Shu, F.H., Adams, F.C. Lizano, S. {\em ARA \& A}, 25:23, 1987.

\bibitem[Spitzer(1968)]{spitzer:proc68}
Spitzer, L. ,F.H., in {\em Nebulae and Interstellar Matter}, 
eds. Middlehurst B. M., Aller L. H., Univ. of Chicago Press p.1 (1968).

\bibitem[Stahler(2000)]{stahler:proc00}
Stahler, S. W., in 
{\em Proc. 33rd ESLAB symp., Star formationfrom the Small to the Large scales}, 
ESA SP-445, Noordwijk, p.133
(2000).

\bibitem[Stone et al.(2000)]{stone:proc00}
Stone, J. M., Gammie, C.F., Balbus, S. A., Hawley, J. F. in 
{\em Protostars and Plantes IV}, eds. Mannings V., Boss, A. P., Russell, 
S. S., Univ. of Arizona Press p.589 
(2000).

\end{thebibliography}
\end{document}